\documentclass[twocolumn]{aa} 
\usepackage{graphicx}
\begin{document}
   \title{Chemical abundances of planetary nebulae towards the galactic anticenter}

   \author{R. D. D. Costa
          \inst{1}
          M. M. M. Uchida
         \inst{1}
          \and
          W. J. Maciel
          \inst{1}
                    }

   \offprints{R. D. D. Costa}

   \institute{Instituto de Astronomia, Geof\'\i sica e Ci\^encias
                 Atmosf\'ericas (IAG), Universidade de S\~ao Paulo - 
                 Rua do Mat\~ao 1226; 05508-900, S\~ao Paulo SP; Brazil\\
                 \email{roberto@astro.iag.usp.br, 
                 monica@astro.iag.usp.br, \\ maciel@astro.iag.usp.br}
                 \thanks{Based on observations made at the European Southern 
                 Observatory (Chile) and Laborat\'orio Nacional de Astrof\'\i sica 
                 (Brazil)}
        }

   \date{Received ; accepted }

   \abstract{In this paper we report new observations and derive chemical
abundances for a sample of 26 planetary nebulae (PN) located in the anticenter
direction. Most of these nebulae are far away objects, located at 
galactocentric distances greater than about 8 kpc, so that they are
particularly useful for the determination of the radial gradients at
large distances from the galactic center. A comparison of the present
results with previously determined abundances suggests that the 
radial abundance gradients flatten out at distances larger than 
about 10 kpc from the center.
   }

   \authorrunning {Costa et al.}
   \titlerunning {Abundances of PN towards the galactic anticenter}

   \maketitle
%

\section{Introduction}

The presence of radial abundance gradients for several element
ratios in the Milky Way is now well established (see Maciel 1997, 2000 
for a revision). Gradients can be derived from different kinds of objects,
such as O,B stars, HII regions  or  planetary nebulae, and seem to be 
a common property of spiral galaxies. In fact, there are reports of 
gradients in many other galaxies apart from the Milky Way from the 
analysis of HII regions (Searle \cite{searle}, Smith \cite{smith}, 
Blair et al. \cite{blair}, Kennicutt \& Garnett \cite{kennicutt}, 
Ferguson et al. \cite{ferguson}).

The study of radial gradients involves one of the most important constraints 
for galactic evolution models, as they can provide at least three crucial 
informations: i) the magnitude of the gradient itself, generally measured 
in dex/kpc; ii) possible variations in the gradient for different 
galactocentric distances; iii) the time variation in the gradients. 

Regarding the space variations of the gradients, presently available
results are somewhat controversial. Some results for the Milky Way
based on stars (Kaufer et al. \cite{kaufer}) and HII regions (Esteban \&
Peimbert \cite{esteban}, V\'\i lchez \& Esteban \cite{vilchez}) suggest 
decreasing gradients for galactocentric distances $R > 6$ kpc. This is 
confirmed by a recent analysis based on planetary nebulae (Maciel \& Quireza 
\cite{mq99}), and is also supported by chemical evolution models by
Chiappini et al. (\cite{cristina}) and  Matteucci (\cite{matteucci}). 
However, no significant variations in the gradient are detected from a sample of
O, B stars by Smartt (\cite{smartt}, and references therein), and a 
flattened gradient near the solar neighbourhood has been suggested by
Andrievsky et al. (\cite{andrievsky} and references therein) on the basis 
of a sample of cepheid variables.

The shapes of the gradients in spiral  galaxies are also subject to
discussion. Some flattening in the outer regions has been proposed
(D\'\i az \cite{diaz}), while some recent results are consistent with an
exponential gradient with a constant slope (Henry \& Howard \cite{henry}).

Planetary nebulae can play an important role in this context (Maciel \&
K\"oppen \cite{mk94}, Peimbert \& Carigi \cite{pc98}), as nebular abundances 
can be accurately derived, even for objects at large heliocentric distances, 
which is particularly useful for objects in the direction of the galactic
anticenter (Maciel \& Quireza \cite{mq99}). Also, these objects can be 
classified according to the masses and ages of their progenitors, so that 
they are also useful in the study of the time variations of the gradients 
(Maciel \& Costa \cite{mc03}, Maciel et al. \cite{mcu03}). The derivation 
of chemical abundances for a homogeneous and wide sample of PN at large 
galactocentric distances, located basically outside the solar circle of 
about 8 kpc, would therefore provide  the tools for the assessment of the 
gradient properties at these large distances.

In this work, we report the chemical abundances for a sample of PN 
towards the galactic anticenter, where most of the objects placed at large 
galactocentric distances are located. They were selected from the 
ESO-Strasbourg Galactic PN Catalog (Acker et al. \cite{acker}), and the sample 
consists of 26 objects, all of them observed, reduced and analysed 
homogeneously.

\section{Observations}

The observations were carried out in two different observatories: at 
ESO/La Silla, and at LNA/Brazil. At ESO we used the 1.52m telescope and 
the Boller \& Chivens Cassegrain spectrograph, in two observational runs: 
dec/1999 and dec/2000. We used grating \#23 and CCD \#38, which provide 
a dispersion of 2.2 \AA/pix, and  spectral coverage of about 5000~\AA. 
The effective wavelength coverage was reduced to about 3500~\AA \ starting
at 4200~\AA, due to a low S/N ratio shortward of this wavelength.
At LNA, in dec/1995, we used the 1.60m telescope, a Boller \& Chivens 
Cassegrain spectrograph and a combination grating+CCD, which gives a 
dispersion of 4.5 \AA/pixel and spectral coverage of about 4500 \AA.
An east-west long slit of 2 arcsecs width was used for all obervations. 
Table 1 gives the logbook of the observations and also the adopted
distances $d$ (kpc) to the Sun and galactocentric distances $R$ (kpc), 
which we will use in Section~4. Flux calibration was secured 
through the observation of at least three spectrophotometric standards in 
each night. Each object was observed at least twice, and line fluxes 
were derived from the average of all good measures for each object.

\begin{table*}
\caption[]{Log of the observations and distances}
\begin{flushleft}
\begin{tabular}{llccllr}
\hline\noalign{\smallskip}
Object & \ \ \ PN G & RA & DEC & \ Place \& Date &\ $d$  & $R$  \\
\noalign{\smallskip}
\hline\noalign{\smallskip}
A 12     & 198.6$-$06.3  &  06 02 23	&     09 39 03 &	ESO - Dec. 1999 &  2.0 &  9.5	\\
A 18     & 216.0$-$00.2  &  06 56 15	&  $-$02 53 12 &	ESO - Dec. 1999 &  1.6 &  8.9	\\
A20      & 214.9$+$07.8  &  07 23 01	&     01 45 37 &	ESO - Dec. 2000 &  2.0 &  9.3	\\
H 3-75   & 193.6$-$09.5  &  05 40 44	&     12 21 16 &	ESO - Dec. 1999 &  2.4 &  9.9	\\
IC 2003  & 161.2$-$14.8  &  03 56 22	&     33 52 30 &	LNA - Dec. 1995 &  4.7 & 12.0	\\
IC 2165  & 221.3$-$12.3  &  06 21 43	&  $-$12 59 10 &	LNA - Dec. 1995 &  2.0 &  9.2	\\
J 320	   & 190.3$-$17.7  &  05 05 34	&     10 42 21 &	LNA - Dec. 1995 &  6.1 & 13.4	\\
J 900	   & 194.2$+$02.5  &  06 25 57	&     17 47 27 &	LNA - Dec. 1995 &  2.8 & 10.3	\\
K 1-7	   & 197.2$-$14.2  &  05 31 48	&     06 56 09 &	LNA - Dec. 1995 &  3.8 & 11.2	\\
K 2-1	   & 173.7$-$05.8  &  05 03 09	&     30 50 03 &	ESO - Dec. 2000 &  1.1 &  8.7	\\
K 3-66   & 167.4$-$09.1  &  04 36 37	&     33 39 30 &	LNA - Dec. 1995 &  6.7 & 14.1	\\
K 3-70   & 184.6$+$00.6  &  05 58 45 	&     25 18 44 &	ESO - Dec. 1999 &      &  	\\
M 1-6	   & 211.2$-$03.5  &  06 35 45	&  $-$00 05 41 &	ESO - Dec. 1999 &  2.6 &  9.9 \\
M 1-7	   & 189.8$+$07.7  &  06 37 21	&     24 00 30 &	ESO - Dec. 1999 &  5.9 & 13.4	\\
M 1-8	   & 210.3$+$01.9  &  06 53 34	&     03 08 24 &	LNA - Dec. 1995 &  3.4 & 10.7	\\
M 1-9	   & 212.0$+$04.3  &  07 05 19	&     02 46 59 &	LNA - Dec. 1995 &  4.9 & 12.0	\\
M 1-13   & 232.4$-$01.8  &  07 21 15	&  $-$18 08 36 &	LNA - Dec. 1995 &  5.3 & 11.6 \\
M 1-14   & 234.9$-$01.4  &  07 27 56	&  $-$20 13 23 &	LNA - Dec. 1995 &  4.0 & 10.4	\\
M 1-16   & 226.7$+$05.6  &  07 37 19	&  $-$09 38 48 &	LNA - Dec. 1995 &  5.5 & 12.0 \\
M 1-17   & 228.8$+$05.3  &  07 40 22	&  $-$11 32 30 &	LNA - Dec. 1995 &  7.4 & 13.6	\\
M 1-18   & 231.4$+$04.3  &  07 42 06	&  $-$14 21 41 &	ESO - Dec. 1999 &  4.4 & 10.9	\\
M 3-2	   & 240.3$-$07.6  &  07 14 50 	&  $-$27 50 20 &	ESO - Dec. 1999 &  8.8 & 14.1	\\
M 3-3	   & 221.7$+$05.3  &  07 26 35	&  $-$05 21 57 &	ESO - Dec. 1999 &  5.7 & 12.4	\\
NGC 1514 & 165.5$-$15.2  &  04 09 22	&     30 46 00 &	ESO - Dec. 2000 &  0.8 &  8.3	\\
SA 2-21  & 238.9$+$07.3  &  08 08 43	&  $-$19 13 59 &	ESO - Dec. 1999 &  3.3 &  9.7	\\
YC 2-5   & 240.3$+$07.0  &  08 10 42	&  $-$20 31 33 &	ESO - Dec. 1999 &  7.7 & 13.2	\\
\noalign{\smallskip}
\hline
\end{tabular}
\end{flushleft}
\end{table*}

Data reduction was performed using the IRAF package, and followed the standard 
procedure for long slit spectroscopy: correction of bias, dark and flat-field, 
extraction of the spectral profile, wavelength calibration and flux calibration. 
Atmospheric extinction was corrected through the mean coefficients for the observatory 
and season. Line fluxes were calculated assuming gaussian profiles, and a gaussian
de-blending routine was used when necessary. Fluxes calculated for all diagnostic
lines are listed in Table~2 (available electronically at the CDS
\footnote{{\tt http://cdsweb.u-strasbg.fr/}}). Interstellar 
reddening was derived from the difference between measured and predicted 
values for the Balmer ratio H$\alpha$/H$\beta$, assuming Case B (Osterbrock 
\cite{osterbrock}), and adopting the interstellar reddening law  by Cardelli 
et al. (\cite{cardelli}),  with R$_V$=3.1. The derived E(B-V) values are  
given in Table 3 for each nebula.

\begin{table*}
\caption[]{Line fluxes (available electronically)}
\begin{flushleft}
\begin{tabular}{lrrrrrrrr}
\hline\noalign{\smallskip}
Ion &   A 12 &   A 18 &   A 20 & H3-75   &   IC 2003  &   IC 2165  &   J 320    &   J 900 \\
\noalign{\smallskip}
\hline\noalign{\smallskip} 
H$\gamma$ 4340	&	54.0	&	34.0	&	27.0	&	68.0	&	26.0	&	53.0	&	63.0	&	59.0  \\	
{\rm[O III]} 4363	&	8.7	&	10.6	&	3.8	&	13.4	&	11.5	&	14.3	&	13.6	&	9.6   \\	
... &  ... &   ... &  ... &  ... & ...   &  ...  &  ...  &  ...   \\
\noalign{\smallskip}
\hline
\end{tabular}
\end{flushleft}
\end{table*}

A preliminary analysis of the objects observed in 1995 was already published (Costa et al. 
\cite {costa97}); however, for the present work we rederived the physical 
parameters and abundances for the whole sample, in order to assure the 
homogeneity of the results.

\section {Physical parameters and abundances}

Electron densities $n_e$ were derived from the [SII]$\lambda$6717/6731 ratio, while 
electron  temperatures $T_e$ were estimated from both the [OIII]$\lambda$4364/5007 and 
[NII]$\lambda$ 5755/6584 line ratios. When both line ratios were available, the
[OIII] temperature was used to derive the ionic concentrarions of higher ionization 
potential ions such as O$^{+2}$, Ar$^{+2}$, Ar$^{+3}$, Ne$^{+2}$, and the [NII] 
temperature for lower ionization potential ions like O$^{+}$, N$^{+}$, S$^{+}$. 
In the cases when only one temperature was available, it was used for all ions.
For the objects A20 and NGC 1514 no temperatures were available, so that an 
average value of 10$^{4}$ K was adopted. Concerning densities, for these 
nebulae the [SII] ratio was not available, and a typical value of 5000 cm$^{-3}$ 
was adopted. Table 3 shows the adopted electron densities (cm$^{-3}$) and
temperatures (K) for all objects.

\begin{table*}
\caption[]{Physical parameters}
\begin{flushleft}
\begin{tabular}{lrrrrr}
\hline\noalign{\smallskip}
Object & $E(B-V)$ & $T_e$ [OIII]  & $T_e$ [NII] & $n_e$[SII]  \\
\noalign{\smallskip}
\hline\noalign{\smallskip}
A 12	    & 0.440 &	11912	&  9144 &	255	\\
A 18	    & 0.969 &	15736	& 14317 &	364	\\
A 20	    & 0.0   &		&	  &  5000	\\
H 3-75    & 0.290 &	12813	&  9013 &	119	\\
IC 2003   & 0.301 &	12080 & 13276 &  5144	\\
IC 2165   & 0.417 &	12519	& 14075 &  5715	\\
J 320	    & 0.248	&     11944	& 12815 &  4915	\\
J 900	    & 0.558 &	11320	& 12513 &  4016	\\
K 1-7	    & 0.144 &	16123	& 11329 &    92	\\
K 2-1	    & 0.433 &	11375	&       &   493	\\
K 3-66    & 0.660 &	 9057	& 13143 &  4725	\\
K 3-70    & 1.530 &	17500	& 11643 &  2361	\\
M 1-6	    & 1.260 &	10558	& 11306 &  8426	\\
M 1-7	    & 0.129 &	10244	& 11016 &  1135	\\
M 1-8	    & 0.374 &	15563	& 10307 &   482	\\
M 1-9	    & 0.540 &	10922	& 10435 &  4516	\\
M 1-13    & 0.208 &	8196	&  9601 &   994	\\
M 1-14    & 0.757 &	12805	&  9987 &  6784	\\
M 1-16    & 0.450 &	10657	& 10351 &  2311	\\
M 1-17    & 0.650 &	10008	& 10525 &  6030	\\
M 1-18    & 0.088 &	11224	& 10896 &    69	\\
M 3-2	    & 0.317 &		& 10164 &	230	\\
M 3-3	    & 0.250 &	13826	&  9867 &	349	\\
NGC 1514  & 0.410 &		&	  &  5000	\\
SA 2-21    & 0.200 &	12720	& 15152 &	351	\\
YC 2-5    & 0.290 &	11477	& 11477 &  5000	\\
\noalign{\smallskip}
\hline
\end{tabular}
\end{flushleft}
\end{table*}

Ionic abundances were calculated using a three-level atom model by solving 
the statistical equilibrium equations, including radiative and collisional 
transitions. Elemental abundances were then derived using ionization correction 
factors to account for unobserved ions of each element. The atomic data and ICFs
used in this work are same as adopted by Escudero \& Costa (\cite{ec01}), to
which the reader is referred for details. The adopted formulae for the ICFs
are given below. 

$$ \frac {\rm{He}} {\rm{H}} = \frac {\rm{He^{+}}} {\rm{H^{+}}} + \frac {\rm{He^{++}}} {\rm{H^{+}}} $$
$$ \frac {\rm{O}} {\rm{H}} = (\frac {\rm{O^{++}}} {\rm{H^{+}}} + \frac {\rm{O^{+}}} {\rm{H^{+}}})
\frac {\rm{He}} {\rm{He^{+}}} $$
$$ \frac {\rm{N}} {\rm{H}} = \frac {\rm{N^{+}}} {\rm{H^{+}}} \frac {\rm{O}} {\rm{O^{+}}} $$
$$ \frac {\rm{S}} {\rm{H}} = (\frac {\rm{S^{+}}} {\rm{H^{+}}} + \frac {\rm{S^{++}}} {\rm{H^{+}}})
(1 - (1- \frac {\rm{O^{+}}} {\rm{O}})^{3})^{-1/3} $$
$$ \frac {\rm{Ar}} {\rm{H}} = 1.34 \frac {\rm{Ar^{++}}} {\rm{H^{+}}}
\frac {\rm{O}} {\rm{O^{++}}} $$
$$ \frac {\rm{Ne}} {\rm{H}} = \frac {\rm{Ne^{++}}} {\rm{H^{+}}} \frac {\rm{O}} {\rm{O^{++}}} $$

\noindent
For the cases where [SIII] 6312 is not available, we adopted the
expression given by Kingsburgh \& Barlow (\cite{kingsburgh}) to estimate 
the S$^{++}$ abundance. As in Escudero \& Costa (\cite{ec01}), we calculated 
helium abundances using the recombination theory, with recombination 
coefficients from P\'equignot et al. (\cite{pequignot}), and correcting 
the HeI abundance from collisional effects using the correction terms 
from Kingdon \& Ferland (\cite{kingdon}). He I  abundances were
determined as weighted averages from the 4471, 5876 and 6678 lines,
using the line fluxes as weights. He II abundances were derived
from the 4686 line. The obtained ionic abundances
are given in Table 4, where \lq\lq E-05\rq\rq\ means $10^{-05}$, etc.

\begin{table*}
\caption[]{Ionic Abundances}
\begin{flushleft}
\begin{tabular}{lrrrrrrrr}
\hline\noalign{\smallskip}
Object  &   He I    &   He II   &   N II    &   S II    &   S III   &   O II    &   O III   &   Ar III  \\
\noalign{\smallskip}
\hline\noalign{\smallskip}
A 12 &   0.111   &   0.008   &   8.45E-05    &   2.82E-06    &   3.23E-06    &   6.21E-04    &   1.69E-04    &   9.03E-07    \\
A 18 &   0.141   &   0.010   &   4.27E-05    &   5.25E-07    &   1.87E-06    &   4.43E-05    &   4.60E-05    &   5.25E-07    \\
A 20 &   0.042   &   0.083   &       &   1.59E-07    &   7.68E-06    &   4.63E-05    &   1.66E-04    &   9.93E-08    \\
H 3-75   &   0.069   &   0.003   &   2.50E-05    &   9.15E-07    &   2.15E-06    &   1.45E-04    &   1.77E-04    &   6.68E-07    \\
IC 2003  &   0.039   &   0.054   &   1.50E-06    &   7.99E-08    &   1.25E-06    &   1.07E-05    &   2.04E-04    &   4.63E-07    \\
IC 2165  &   0.047   &   0.029   &   2.51E-06    &   9.38E-08    &   1.75E-06    &   9.28E-06    &   2.09E-04    &   4.98E-07    \\
J 320    &   0.101   &   0.006   &   2.25E-07    &   2.33E-08    &   9.80E-07    &   4.23E-06    &   2.57E-04    &   4.06E-07    \\
J 900    &   0.056   &   0.038   &   4.22E-06    &   1.44E-07    &   1.33E-06    &   2.96E-05    &   2.47E-04    &   4.75E-07    \\
K 1-7    &   0.100   &   0.016   &   2.79E-05    &   1.07E-06    &   1.11E-06    &   1.16E-04    &   8.60E-05    &   5.33E-07    \\
K 2-1    &   0.056   &   0.066   &   1.89E-05    &   3.25E-07    &   2.41E-06    &   1.63E-04    &   2.52E-04    &   1.02E-06    \\
K 3-66   &   0.083   &       &   9.26E-06    &   8.93E-08    &   2.03E-06    &   4.83E-05    &   1.21E-04    &   6.86E-07    \\
K 3-70   &   0.138   &   0.023   &   1.16E-04    &   9.61E-07    &   2.14E-06    &   1.05E-04    &   7.30E-05    &   7.09E-07    \\
M 1-6    &   0.081   &       &   2.42E-05    &   1.91E-07    &   1.73E-06    &   1.88E-04    &   3.12E-05    &   4.34E-07    \\
M 1-7    &   0.132   &   0.005   &   5.97E-05    &   9.79E-07    &   5.58E-06    &   2.26E-04    &   3.85E-04    &   3.45E-06    \\
M 1-8    &   0.122   &   0.034   &   1.02E-04    &   1.20E-06    &   1.27E-06    &   2.56E-04    &   1.28E-04    &   1.10E-06    \\
M 1-9    &   0.062   &   0.001   &   1.12E-05    &   1.36E-07    &   1.07E-06    &   8.93E-05    &   1.18E-04    &   4.17E-07    \\
M 1-13   &   0.122   &   0.012   &   1.32E-04    &   1.23E-06    &   5.56E-06    &   4.80E-04    &   7.79E-04    &   4.74E-06    \\
M 1-14   &   0.086   &       &   1.59E-05    &   2.23E-07    &   5.95E-07    &   1.53E-04    &   4.79E-05    &   3.64E-07    \\
M 1-16   &   0.104   &   0.023   &   1.40E-04    &   3.42E-07    &   1.18E-06    &   2.40E-04    &   3.46E-04    &   1.45E-06    \\
M 1-17   &   0.069   &   0.011   &   3.86E-05    &   1.50E-06    &   6.29E-06    &   1.61E-04    &   4.67E-04    &   1.57E-06    \\
M 1-18   &   0.136   &   0.017   &   9.21E-05    &   1.14E-06    &   5.97E-06    &   1.75E-04    &   1.91E-04    &   1.38E-06    \\
M 3-2    &   0.134   &   0.077   &   1.80E-04    &   1.21E-06    &   6.03E-07    &   8.95E-05    &   1.74E-05    &   2.51E-07    \\
M 3-3    &   0.100   &   0.029   &   1.79E-04    &   3.00E-07    &   1.22E-04    &   1.51E-04    &   1.22E-04    &   8.38E-07    \\
NGC 1514 &   0.076   &   0.005   &       &   1.63E-07    &   3.40E-06    &   1.96E-04    &   4.15E-04    &   4.48E-07    \\
SA 2-21  &   0.110   &   0.019   &   1.67E-05    &   5.72E-07    &   2.38E-06    &   2.99E-05    &   1.44E-04    &   9.65E-07    \\
YC 2-5  &   0.017   &   0.054   &       &       &       &       &   2.66E-04    &   1.74E-07    \\
\noalign{\smallskip}
\hline
\end{tabular}
\end{flushleft}
\end{table*}

\subsection{Abundances and distance-independent correlations}

Derived elemental abundances are given in table 5. The helium abundance (He/H) 
is relative to hydrogen, and the remaining elemental abundances are given as 
$\epsilon$(X)=log(X/H)+12. The N/O ratio is also given, as this ratio
is useful in order to classify the nebulae, and is also related to the
progenitor star mass. We have also examined the classification of the  
objects in our sample according to the classification scheme proposed by 
Peimbert (\cite{peimbert}), which relies basically on elemental abundances. 
All nebulae in the sample are disk objects, belonging to types I, II and III. 
Subtypes IIa and IIb (Fa\'undez-Abans \& Maciel \cite{faundez}) have also
been included. Bulge (type V) or halo (type IV) objects were avoided 
(see the last column of Table 5). 

Average helium abundances in our sample are He/H = 0.160 for type I PN and
He/H = 0.107 for type II objects. For oxygen, we obtain $\epsilon$(O) =
8.42 and  $\epsilon$(O) = 8.66, respectively, thus showing some 
indication of ON cycling in type I PN, presumably those with more
massive progenitors in which the second dredge up process has 
operated (see for example de Freitas Pacheco \cite{pacheco},
Kingsburgh \& Barlow \cite{kingsburgh}).

\begin{table*}
\caption[]{Abundances}
\begin{flushleft}
\begin{tabular}{lrrrrrrrrrr}
\hline\noalign{\smallskip}
Object . & He/H & $\epsilon$(O) & $\epsilon$(N) & 
$\epsilon$(S) & $\epsilon$(Ar)& log(N/O) & Type\\
\noalign{\smallskip}
\hline\noalign{\smallskip}
A 12	   &	0.119	&  8.93 &	8.06	&  6.78  &	6.78	&  $-$0.87  &	IIa	\\
A 18	   &	0.152	&  7.99 &	7.97	&  6.40  &	6.17	&  $-$0.02  &	I	\\
A 20	   &	0.125	&  8.80 &		&  6.66  &	6.86	&           &	IIa	\\
H 3-75   &	0.071	&  8.52 &	7.76	&  6.52  &	6.23	&  $-$0.76  &	IIb	\\
IC 2003  &	0.094	&  8.71 &	7.86	&  6.53  &	6.19	&  $-$0.85  &	IIb	\\
IC 2165  &	0.076	&  8.55 &	7.98	&  6.64  &	6.05	&  $-$0.57  &	IIb	\\
J 320	   &	0.107	&  8.44 &	7.17	&  6.45  &	5.77	&  $-$1.27  &	IIb	\\
J 900	   &	0.094	&  8.67 &	7.82	&  6.42  &	6.08	&  $-$0.85  &	IIb	\\
K 1-7	   &	0.115	&  8.37 &	7.75	&  6.36  &	6.29	&  $-$0.62  &	IIb	\\
K 2-1	   &	0.122	&  9.19 &	8.26	&  6.62  &	6.28	&  $-$0.93  &	IIa	\\
K 3-66   &	0.083	&  8.23 &	7.51	&  6.39  &	6.11	&  $-$0.72  &	IIb	\\
K 3-70   &	0.161	&  8.32 &	8.36	&  6.51  &	6.43	&     0.04  &	I	\\
M 1-6	   &	0.081	&  8.34 &	7.45	&  6.28  &	6.61	&  $-$0.89  &	IIb	\\
M 1-7	   &	0.137	&  8.80 &	8.23	&  6.86  &	6.88	&  $-$0.57  &	IIa	\\
M 1-8	   &	0.156	&  8.69 &	8.29	&  6.41  &	6.75	&  $-$0.40  &	IIa	\\
M 1-9	   &	0.063	&  8.32 &	7.42	&  6.11  &	6.00	&  $-$0.90  &	IIb	\\
M 1-13   &	0.134	&  9.14 &	8.58	&  6.88  &	7.05	&  $-$0.56  &	IIa	\\
M 1-14   &	0.086	&  8.30 &	7.32	&  5.91  &	6.31	&  $-$0.98  &	III	\\
M 1-16   &	0.127	&  8.86 &	8.62	&  6.23  &	6.60	&  $-$0.24  &	IIa	\\
M 1-17   &	0.081	&  8.86 &	8.24	&  6.98  &	6.52	&  $-$0.62  &	IIb	\\
M 1-18   &	0.153	&  8.62 &	8.34	&  6.88  &	6.60	&  $-$0.28  &	I	\\
M 3-2	   &	0.212	&  8.23 &	8.53	&  6.27  &	6.51	&   0.30  &	I	\\
M 3-3	   &	0.129	&  8.54 &	8.62	&  6.49  &	6.51	&   0.08  &	I	\\
NGC 1514 &	0.081	&  8.81 &		&  6.61  &	5.97	&   	    &	IIb	\\
SA 2-21  &	0.129	&  8.31 &	8.05	&  6.61  &	6.26	&  $-$0.22  &	IIa	\\
YC 2-5   &	0.071	&  9.05 &		&        &	6.00	&	    &	IIb	\\
\noalign{\smallskip}
\hline
\end{tabular}
\end{flushleft}
\end{table*}

\begin{figure*}
   \centering
   \includegraphics[angle=-90,width=12.0cm]{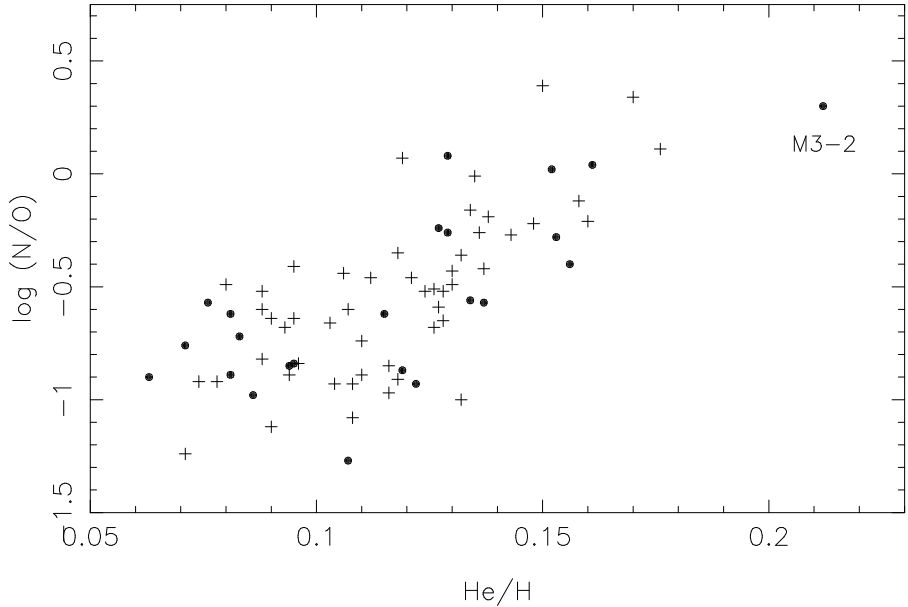}
      \caption{$\log$(N/O) as a function of the helium abundance. 
      Our new results are presented as solid dots, while the 
      crosses show the abundances derived in our previous
      work, as quoted in the text.}
         \label{fig1}
   \end{figure*}

The relative abundances can be better examined using distance-independent correlations,
as can be seen in Figure \ref{fig1}, which shows $\log$(N/O)$ \times$ He/H.
Our new data are presented as solid dots, and the crosses show the results of
our previous work as given by Costa et al. (\cite{costa96}), de Freitas Pacheco 
et al. (\cite{pacheco89}, \cite{pacheco91}, \cite{pacheco92}), and
Maciel et al. (\cite{maciel90}). As can be seen in the previous papers,
all objects have been observed and analyzed in a homogeneous way,
so that the total sample includes about 80 nebulae in the disk alone,
forming one of the largest database of accurate abundances of planetary
nebulae available. This is important, since distance independent correlations
as shown by Fig.~\ref{fig1} are much better defined when a large sample
is considered. It can be seen that our new data fit rather nicely
in the observed correlation.

The plot shows the expected trend of an increase in N/O as He/H increases, reflecting 
the enrichment of He and N due to dredge-up episodes, as predicted by stellar evolution 
models. Some N/O enrichment can also be observed by comparing the data in Fig. \ref{fig1}
(or Table~5) with the corresponding values of the N/O ratio for HII regions of 
similar metallicity, measured by the O/H ratio, as given for example by Henry \&
Worthey (\cite{hw99}).

\begin{figure*}
   \centering
   \includegraphics[angle=-90,width=12.0cm]{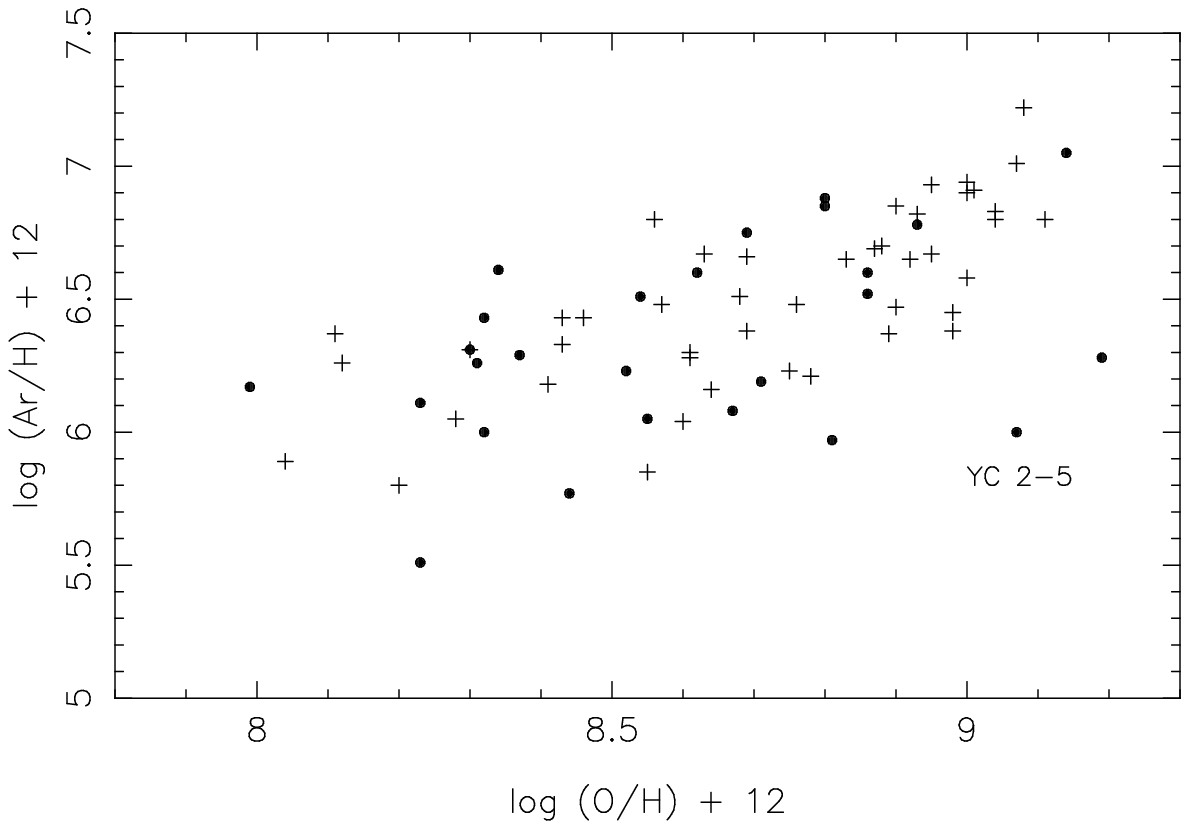}
      \caption{Argon abundances as a function of the O/H ratio.
      The symbols are as in Fig. 1.}
         \label{fig2}
   \end{figure*}

\begin{figure*}
   \centering
   \includegraphics[angle=-90,width=12.0cm]{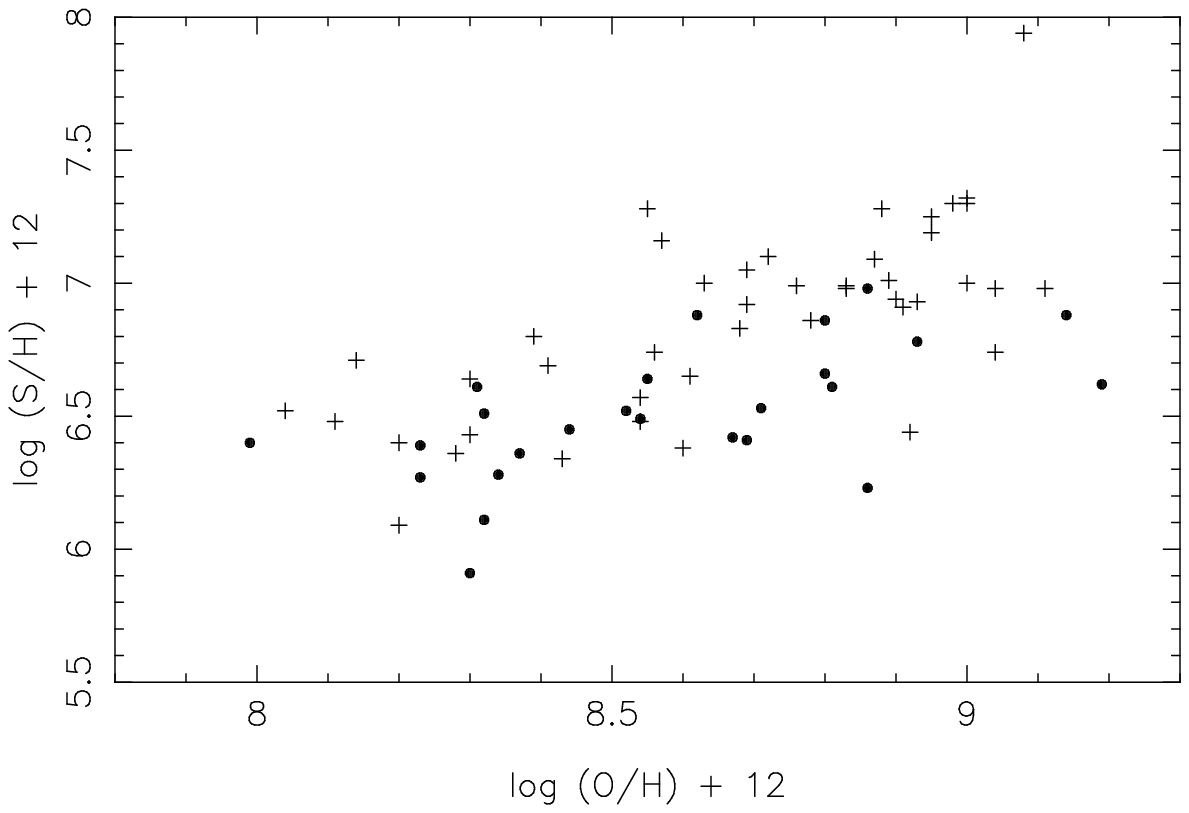}
      \caption{Sulfur abundances as a function of the O/H ratio.
      The symbols are as in Fig. 1.}
         \label{fig3}
   \end{figure*}

Figures \ref{fig2} and \ref{fig3} show the correlations with the O/H ratio 
of the Ar/H and S/H abundances, respectively. These correlations are also 
well defined, and reflect the mass distribution of PN progenitors, as younger, 
more massive objects are expected to form from an interstellar medium 
richer in alpha-elements. The Ar abundance of the nebula YC 2-5 (Sa 2-22) 
shows a larger deviation from the trend as compared to the remaining 
objects, as indicated in Figure \ref{fig2}. This object shows a normal 
He abundance, and its oxygen abundance can also be considered as normal, 
albeit larger than the average. If that is true, the argon abundance is 
probably a lower limit, which may be due to the fact that its electron 
density is not accurately determined.

\subsection{Comparison with previous results from the literature}

Most of the nebulae studied in this work are objects located at relatively 
large distances from the sun, for which accurate elemental abundances 
are being determined for the first time. Some objects, however, have 
been studied by other groups, so that it is interesting to compare 
our new results with previously determined abundances in the literature.

Considering first the earlier work, the objects IC~2003, IC~2165, J320 
and J900 are part of the detailed study by Aller \& Czyzak (\cite{aller}) 
based on image-tube scanner measurements and making use of theoretical models
to represent individual line intensities. Regarding the He/H abundances, 
our data agree with their results within 4\% for IC~2003, J320 and J900, 
with a larger discrepancy of about 38\% for IC~2165, for which the 
abundance by Aller \& Czyzak (\cite{aller}) is higher than ours. Our electron
temperatures for this object do not differ from the adopted values by
Aller \& Czyzak (\cite{aller}) and more recently by Kholtygin (\cite{kholtygin})
by more than 8\%. However, a larger discrepancy in the electron 
temperatures of 11 to 16\% and in the helium abundance again of
38\% is apparent for this nebula from the detailed work by Hyung 
(\cite{hyung}), so that our ionization correction factor may not account 
completely for the total abundance, and our result is probably underestimated. 
Considering now the heavy elements O, N, S and Ar, for IC~2003, IC~2165 
and J900 the agreement is good, with an average discrepancy of a factor 1.3. 
The worst case is the Ar abundance of IC~2165, reaching about a factor 1.8, 
which is probably a result of the weaker lines of this element, or may be due
the the adopted electron temperature. For these elements the difference
of our results to the analysis by Hyung (\cite{hyung}) is of a factor~1.4
in average. Note however that our oxygen abundance
agrees very well with the more recent work of Kholtygin (\cite{kholtygin}),
as we will see later on this section. For J320 the agreement is very good for 
oxygen, which is well measured in both works, and poorer for the other elements.
It should be noted that the sulfur and nitrogen abundances in 
the work of Aller \& Czyzak (\cite{aller}) are of quality C or D, and that
some recent work that we will quote later also produces discrepant results,
which may be an indication of some variation inside the nebula.

In the spectrophotometric surveys performed by K\"oppen et al. (\cite{koeppen}) 
and Cuisinier et al. (\cite{cuisinier}), about 140 planetary nebulae have 
been studied on the basis of good quality spectra in which plasma diagnostics 
have been tested with theoretical photoionization models. The first survey has 
four objects in common with our work, namely J320, J900, M1-6 and M3-2. The 
He/H abundances for these objects are in agreement with our results within an 
average of 26\%. The worst cases are J320 and M1-6. In the former, the
helium abundance was probably underestimated by K\"oppen et al. (\cite{koeppen}),
as our value agrees very well with Aller \& Czyzak (\cite{aller}) and with
the recent work of Milingo et al. (\cite{milingo}), as we will see later on
in this section. For M1-6, the helium abundance obtained by
K\"oppen et al. (\cite{koeppen}) is reportedly less accurate, due to 
incompleteness of the ionization correction. Regarding the heavy elements, 
the average discrepancy between our results and those by K\"oppen et al. 
(\cite{koeppen}) is a factor 1.3, which means excellent agreement, for all 
elements and objects well measured by them, that is, those not marked with a 
colon (:). The remaining elements, for which their abundances are reportedly  
less accurate, present a much larger deviation, so that we believe our 
measurements do represent an improvement in the knowledge of these objects.

In the second study by Cuisinier et al. (\cite{cuisinier}), featuring 
nebulae located above the galactic plane, chemical abundances have been 
derived  for the objects M1-9, M1-16 and YC~2-5. Considering the first 
two, the He/H abundances agree with our values within 20\%, and the 
heavy elements show an average discrepancy of a factor 1.8, which is
especially large for argon. For YC~2-5 
the discrepancy is higher, about a factor 2.5, but it should be mentioned
that Cuisinier et al. (\cite{cuisinier}) could not derive the sulfur 
abundance and their argon abundance is uncertain due to incompleteness 
of the ionization correction. On the other hand, we could not measure
the nitrogen abundance, which is indeed very low according to Cuisinier 
et al. (\cite{cuisinier}). Also, its odd postion in Fig.~2 could be
conciliated if its oxygen abundance is lower, as suggested by Cuisinier
et al. (\cite{cuisinier}). Clearly, this object deserves further study.

The comprehensive work of Kingsburgh \& Barlow (\cite{kingsburgh}, 
hereafter KB94), based on optical spectrophotometric observations 
and UV data from the IUE satellite. Five nebulae from Table 1 
(IC 2003, J900, M1-8, M1-13, and M3-3) are included in the KS94 sample, 
for which a generally good agreement is observed in the abundances 
within the uncertainties of each work. This is particularly interesting,
as the abundances derived by KB94 use independent ICFs, which are based 
on detailed photoionization models. For the He/H ratio, the abundances 
of these objects derived by KS94 show an average deviation of 
$\Delta$(He/H) = 0.018 relative to our data, which corresponds roughly 
to 15\%. The worst case is J900, for which a discrepancy of about 30\% 
is obtained, but as we will see later, our value is in very good 
agreement with the recent result by Kwitter et al. (\cite{kwitter}). 
For the heavy elements, the average deviations (dex) of the 
KS94 sample relative to our data are $\Delta\epsilon$(O) = 0.14, 
$\Delta\epsilon$(N) = 0.15, $\Delta\epsilon$(Ar) = 0.27, and 
$\Delta\epsilon$(S) = 0.32, which reflects the fact that the sulfur and 
argon abundances are generally less well known  than the oxygen and 
nitrogen data, in view of their lower relative abundances. The average 
discrepancy factors for oxygen and nitrogen are about 1.5,
which are quite reasonable. For argon the discrepancy factor is somewhat
larger, roughly 2.2. For sulfur we get a factor of 1.4 for the
objects IC~2003, M1-8 and M1-13. For J~900 and M3-3 the agreement 
of the results by KB94 and our data is poor, but the sulfur 
abundances of these objects were not well determined in the work of KB94.

Perinotto \& Corradi (\cite{perinotto}) analyzed the chemical structure 
of some bipolar planetary nebulae by means of long slit spectrophotometry, 
considering several regions inside the nebulae in order to detect
spatial abundance variations through the nebulae. There are three objects 
in common with our work, namely M1-13, M1-16 and M3-2. The He/H abundances 
for these objects are similar to our values within 20\%, especially the 
relatively large abundance found in M3-2, which is close to our result
and also to the data of K\"oppen et al. (\cite{koeppen}), showing
that this object is really helium-rich, as can be seen in Figure~1. 
The abundances of the heavy elements O, N, S and Ar are in agreement
with our values within a factor of 2, in average, with a better agreement 
for M1-16 and M3-2 and a poorer one for M1-13, in which case our results
are closer to those by KB94.

Kholtygin (\cite{kholtygin}) determined oxygen abundances of about 70 
planetary nebulae taking into account the effect of small temperature 
and density fluctuations on the line intensities of the nebular lines. 
The objects IC~2003, IC~2165, J320 and J900 are included in his sample, 
for which the oxygen abundances are $\epsilon$(O) = 8.70, 8.50, 8.58 
and 8.45, respectively. The agreeement with our results is excellent, 
within 10\% for the first three nebulae and of a factor 1.7 for J900. 
The case of IC~2165 is particularly interesting, as we have mentioned
the discrepancy of the helium abundance and electron temperature relative to 
some previous work.

The objects J320 and J900 were part of a recent survey of PN which aimed 
at providing a homogeneous spectroscopic database in the wavelength range 
of 3600--9600~\AA \ (Milingo et al. \cite{milingo}, Kwitter et al. \cite{kwitter}).
This survey had the advantage of using the near infrared lines of [SIII] 
$\lambda$ 9069, 9532 \AA, so that their abundances are expected to be more 
accurate, especially for sulfur, in view of a better determination of the
ionic abundances of S$^{+2}$. A comparison of our results for J900 
(Kwitter et al. \cite{kwitter}) shows a very good agreement for all elements 
better than 0.2 dex, or with an average discrepancy of a factor 1.3. 
This fact supports our data for this nebula in view of the discrepancies
found relative to some of the earlier work already quoted, especially
for nitrogen (Aller et al. \cite{aller}), argon (K\"oppen et al.
\cite{koeppen}), oxygen (Kholtygin \cite{kholtygin}), and the elements
argon and sulfur (Kingsburgh \& Barlow \cite{kingsburgh}).
For J320 (Milingo et al. \cite{milingo}) the agreement is good for oxygen 
and argon, with an average discrepancy of a factor 1.3, that is, the
abundance differences are within 0.2 dex. For sulfur, we find an
abundance higher by a factor 2, or $\Delta\epsilon \simeq 0.30$ dex. In
fact, both our results for the sulfur abundances of J320 and J900 confirm
the remarks by Kwitter et al. (\cite{kwitter}) and Milingo et al. 
(\cite{milingo}), in the sense that use of the optical lines only probably 
leads to an overestimate of the sulfur abundance. For nitrogen, there is
no agreement as this element is probably underestimated in our sample.

\section {The radial abundance gradient}

One of the most interesting consequences of the determination of
abundances of PN located in the galactic anticenter is the
effect they might have on the determination of the space variations
in the radial abundance gradients. An analysis by Maciel \& Quireza
(\cite{mq99}, hereafter MQ99) suggested that the gradients probably 
flatten out at galactocentric distances larger than $R \simeq 10$ kpc 
(assuming $R_0 = 7.6$ for the galactocentric distance of the LSR), especially
for the O/H and Ne/H ratios. It is therefore interesting to investigate
whether the presently derived abundances follow the same trend.

We have then produced a merged sample including the objects of MQ99 
and the present results. Both samples form a homogeneous group, as 
the abundances in MQ99 were obtained using basically the same methods 
as here. In fact, all abundances are generally similar within the average 
uncertainty of about 0.1 to 0.2 dex. 

\begin{figure*}
   \centering
   \includegraphics[angle=-90,width=12.0cm]{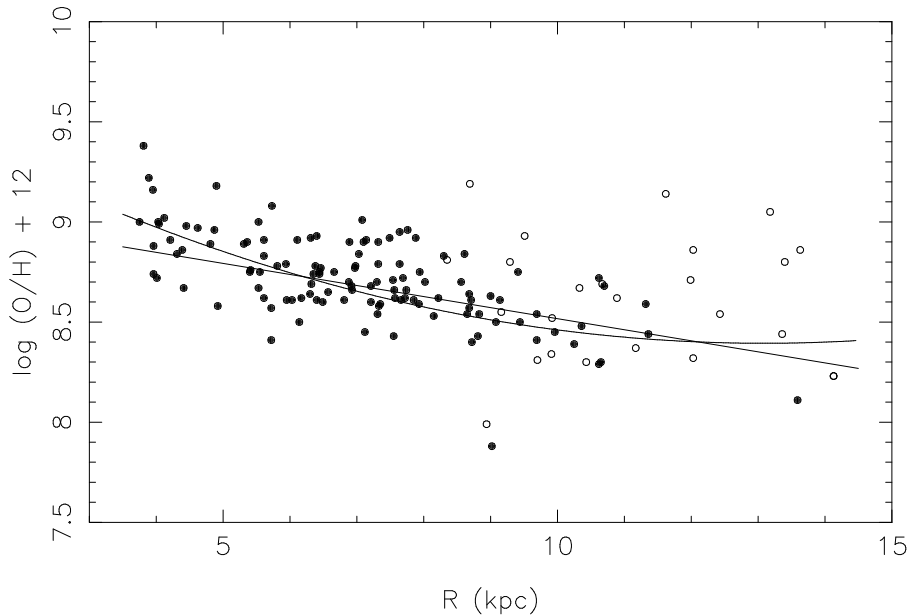}
      \caption{The O/H gradient from the merged sample of Maciel
    \& Quireza (\cite{mq99}, filled circles) and the present 
      sample (open circles). Linear and quadratic least squares
     fits are also shown.}
         \label{fig4}
   \end{figure*}

In Figure \ref{fig4} we plot our new results for oxygen (open circles) along with 
the nebulae presented by MQ99 (filled circles). We have not included the 
object K3-70, for which apparently no reliable distance has been determined.
We adopted the same galactocentric distance of the LSR as MQ99 
and the PN distances are from Cahn et al. (\cite{cahn}), except for H3-75, 
for which we have used the distance given by Amnuel et al. (\cite{amnuel}).
The adopted heliocentric distances $d$ and the galactocentric distances $R$ 
(kpc) are shown in the last two columns of Table~1. 
Use of a different distance scale such as the catalogue by Maciel (\cite{m84}) 
produces essentially the same results. As has been shown before (see for example 
Maciel \& K\"oppen \cite{mk94}, hereafter MK94), the use of different 
statistical distances does not significantly affect the magnitudes 
of the gradients. 

The results indicate that the radial abundance gradient, which 
is reasonably  constant between galactocentric distances from 4 to 10 kpc, 
decreases for greater distances. The effect of the anticenter nebulae 
of the present work in the determination of the gradient is clearly seen 
in Figure \ref{fig4}, as our sample was chosen specifically to cover galactocentric 
distances greater than $R_0$, so they fill the lack of data for these distances
in previous samples, such as MQ99 and MK94.  This can also be seen 
by the least squares fits shown, both for a linear correlation (the straight
line in Figure \ref{fig4}) and a quadratic regression (curve). For the latter,
the inner gradient is about $-0.09$ dex/kpc for galactocentric radii
between 4 and 5 kpc, reaching an essentially flat gradient for 
$R \geq 11$ kpc. The average slope as indicated by the straight line 
is $-0.05$ dex/kpc, which is slightly lower than previous determinations 
for galactocentric distances close to $R_0.$ The flattening at large
distances is seen by the difference between the straight line and
the parabola, especially at $R < 5$ kpc and $R > 12$ kpc.
This result supports other observational results for HII regions (see for example
V\'\i lchez \& Esteban (\cite{vilchez}) and some theoretical models for 
the evolution of spiral galaxies (see Pagel \cite{pagel} and references therein).   

It should be noted that the dispersion of the abundances 
increases at large galactocentric distances, probably
due to the fact that the nebulae are located farther away than solar
neighbourhood objetcs, which are better studied. For the remaining elements
a similar behaviour is observed, which is also seen in comparison with
the earlier sample of Maciel \& K\"oppen (\cite{mk94}). 
For a detailed discussion of these effects the reader is referred to the 
recent work by Maciel \& Costa (\cite{mc03}) and Maciel et al. (\cite{mcu03}).

It is interesting to compare the variation of the O/H gradient with 
galactocentric radius as indicated by Figure \ref{fig4} with some recent results
derived from different galactic objects. Data from O, B stars as discussed
by Smartt (\cite{smartt}, see also Rolleston et al. \cite{rolleston}) 
suggest a constant slope for all galactocentric radii. The different
behaviour of both samples is probably related to the time variation
of the O/H gradient, as recently discussed by Maciel et al. (\cite{mcu03}).
According to that work, the {\it average} gradient, which can be
roughly defined as the gradient from galactocentric radii of 4 kpc 
to about 10 kpc,  has flattened out from $-0.08$ dex/kpc to about $-0.06$ dex/kpc
in the last 5 Gyr, allowing for an average uncertainty of about 0.2 dex/kpc. 
Therefore, it is tempting to conclude that at the time when the oldest disk PN
have been formed -- several Gyr in the past -- the {\it average} gradient
was somewhat steeper than indicated by younger objects such as O, B
stars and HII regions. However, some flattening was already present
near the outer edge of the Galaxy, probably reflecting the lower
star formation rate in that region as compared with the regions closer
to the galactic center. Such flattening is apparent from our PN data,
as can be seen in Figure \ref{fig4}. As the galactic evolution proceeded, the
flattening initially located near the outer Galaxy spread towards
the more central regions, which explains the lower and approximately
constant gradient shown by the O, B stars. Since the temporal 
flattening rate has decreased in the last few Gyr, as suggested by
the results of Maciel et al. (\cite{mcu03}), the gradients presented
by young objects such as O, B stars and HII regions are expected to
be similar, which is supported by the observations, again within an
average uncertainty of 0.02 dex/kpc. Such a scenario is also supported
by recent work on HII regions, for which gradients as low as about
$-0.04$ dex/kpc have been proposed (Deharveng et al.
\cite{deharveng}). On the theoretical side, gradients that flatten out
in time with a decreasing flattening rate in the last few Gyr are 
supported by the recent models proposed by Hou et al. (\cite{hou}).

\begin{acknowledgements}
      This work was partly supported by the brazilian agencies FAPESP and 
      CNPq. M.M.M.U. acknowledges FAPESP for her undergraduate 
      fellowship (Process 00/02813-9).  Observations at ESO/Chile were possible 
      through the FAPESP grant 98/10138-8.
      
\end{acknowledgements}

\end{document}